\documentclass[%
superscriptaddress,
preprint,
 amsmath,amssymb,
 aps,
]{revtex4-2}

\usepackage{graphicx}
\usepackage{dcolumn}
\usepackage{siunitx}

\usepackage{bm}

\usepackage{notes2bib}
\bibnotesetup{
note-name = ,
use-sort-key = false,
placement = mixed
}

\begin{document}

\preprint{}

\title{Excitation of whistler-mode waves by an electron temperature anisotropy in a laboratory plasma
}

\author{Donglai Ma}
\affiliation{Department of Earth, Planetary, and Space Sciences, University of California, Los Angeles, California 90095, USA}

\author{Xin An}
\affiliation{Department of Earth, Planetary, and Space Sciences, University of California, Los Angeles, California 90095, USA}
\author{Jia Han}
\affiliation{Department of Physics, University of California, Los Angeles, California 90095, USA}
\author{Shreekrishna Tripathi}
\affiliation{Department of Physics, University of California, Los Angeles, California 90095, USA}

\author{Jacob Bortnik}
\affiliation{Department of Atmospheric and Oceanic Sciences, University of California, Los Angeles, California 90095, USA}

\author{Anton V. Artemyev}
\affiliation{Department of Earth, Planetary, and Space Sciences, University of California, Los Angeles, California 90095, USA}

\author{Vassilis Angelopoulos}
\affiliation{Department of Earth, Planetary, and Space Sciences, University of California, Los Angeles, California 90095, USA}

\author{Walter Gekelman}
\affiliation{Department of Physics, University of California, Los Angeles, California 90095, USA}
\author{Patrick Pribyl}
\affiliation{Department of Physics, University of California, Los Angeles, California 90095, USA}
\date{\today}

\begin{abstract}
Naturally-occurring whistler-mode waves in near-Earth space play a crucial role in accelerating electrons to relativistic energies and scattering them in pitch angle, driving their precipitation into Earth's atmosphere. Here, we report on the results of a controlled laboratory experiment focusing on the excitation of whistler waves via temperature anisotropy instabilities--the same mechanism responsible for their generation in space. In our experiments, anisotropic energetic electrons, produced by perpendicularly propagating microwaves at the equator of a magnetic mirror, provide the free energy for whistler excitation. The observed whistler waves exhibit a distinct periodic excitation pattern, analogous to naturally occurring whistler emissions in space. Particle-in-cell simulations reveal that this periodicity arises from a self-regulating process: whistler-induced pitch-angle scattering rapidly relaxes the electron anisotropy, which subsequently rebuilds due to continuous energy injection and further excites wave. Our results have direct implications for understanding the process and characteristics of whistler emissions in near-Earth space.

\end{abstract}



\maketitle
Whistler-mode waves are naturally occurring electromagnetic emissions widely found in planetary magnetospheres and many other plasma systems. Propagating below the electron gyrofrequency, these waves play a crucial role in electron dynamics across multiple contexts. In tokamaks, runaway electrons can drive whistler waves unstable, leading to pitch-angle scattering that mitigates these runaways \cite{spong2018first,liu2018role,breizman2019physics}. In the solar wind, whistler waves regulate the nonthermal features of the electron velocity distribution, such as strahl electrons, thereby influencing energy transport and solar wind energetics \cite{hollweg1974electron,gary1999electron,bale2013electron,verscharen2022electron,coburn2024regulation}. In Earth’s magnetosphere, whistler-mode chorus waves serve as a key mechanism for accelerating radiation belt electrons \cite{horne2005wave,thorne2013rapid} and scattering them into the loss cone \cite{ni2008resonant,nishimura2010identifying,kandar2023repetition}, contributing to auroral precipitation.

Whistler-mode waves are introduced into laboratory plasmas either by direct antenna injection \cite{gekelman2011using,stenzel2014magnetic,van2014direct,stenzel2016comparison} or excited through velocity-space instabilities, such as beam \cite{stenzel1977observation,krafft1994whistler,starodubtsev1999resonant,van2015excitation,an2016resonant}, loss-cone \cite{ikegami1968characteristic,booske1985experiments}, and temperature anisotropy distributions \cite{garner1987warm,shalashov2018observation}. Of particular importance, whistler-mode chorus waves arise from temperature anisotropic electrons with $T_{\perp} / T_{\parallel} > 1$ injected from the plasma sheet into the inner magnetosphere ($T_{\perp}$ and $T_{\parallel}$ being the perpendicular and parallel electron temperatures to the background magnetic field, respectively) \cite[e.g.,][]{kennel1966limit,hwang2007statistical,schriver2010generation,santolik2010wave,li2010themis,liu2011excitation,liu2025field}. Despite many spacecraft observations of whistler wave excitation in near-Earth space, there have been far too few laboratory studies of this important process.

\begin{figure}[b]
\includegraphics[width = 0.8\textwidth]{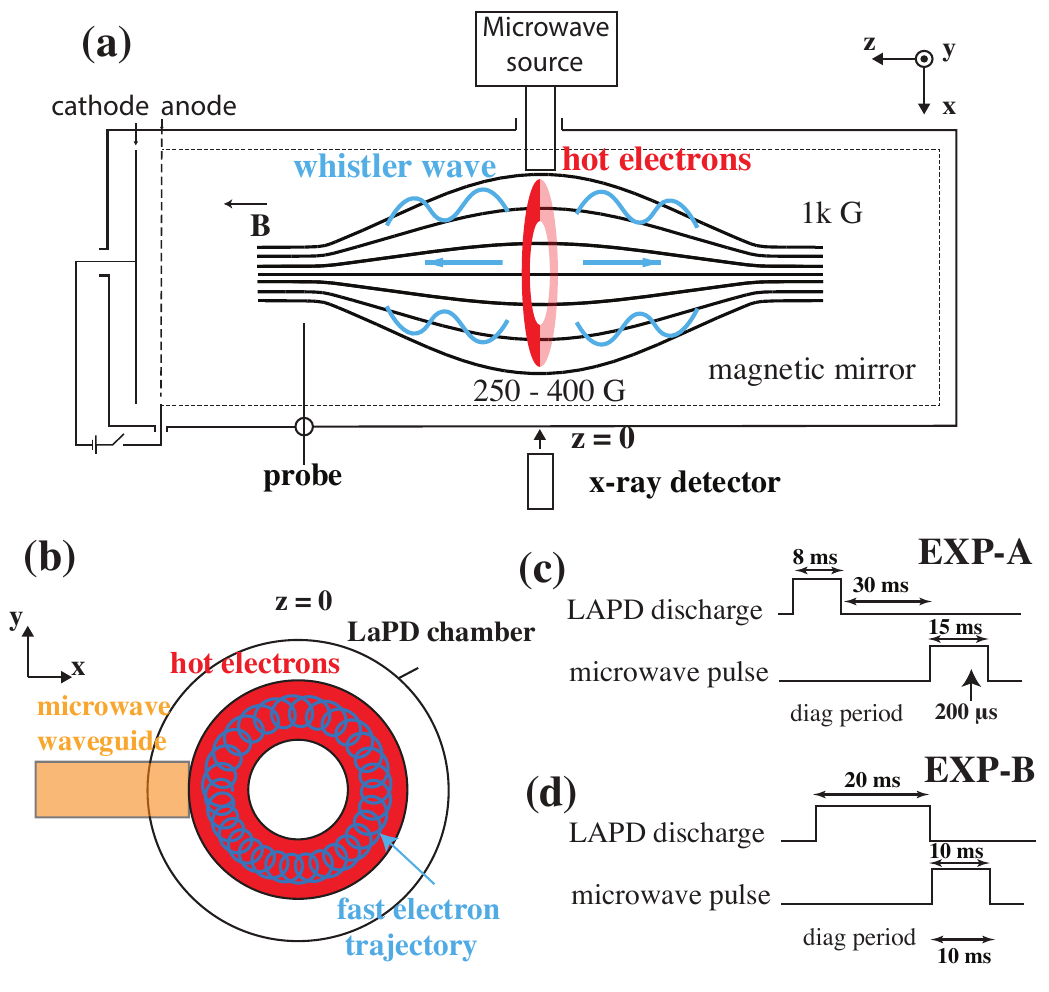}%
\caption{\label{fig:figure1} (a) Schematic side view of the experimental setup (not to scale). The mirror section is about $3.5$\,m. The reference location $z=0$ is at the equator of the magnetic mirror. The B-dot probe is positioned outside of the mirror at $z = 3.2$\,m. (b) Schematic view of the $z = 0$ plane. (c) Two Experimental timing schemes: In EXP-A, each shot lasts $200$ $\mathrm{\mu s}$ with measurements taken at $3$, $5$, $7$, $9$, and $13$\,ms after microwave pulse initiation. In EXP-B, the diagnostic operates continuously for $10$\,ms.}
\end{figure}

In this Letter, we demonstrate the controlled excitation of whistler-mode waves via temperature anisotropy instabilities in a set of stable and highly reproducible experiments, capturing key features of their periodic dynamics in parameter regimes characteristic of near-Earth space. First, our experiments generate whistler waves at a fraction of the electron gyrofrequency using temperature-anisotropic electrons in a magnetic mirror, with dimensionless parameters closely mimicking those in the inner magnetosphere. Second, similar to the continuous injection of energetic electrons from the plasma sheet into the inner magnetosphere during geomagnetic activity, our experiments sustain anisotropic electron populations through continuous microwave heating. These driven systems exhibit a self-regulating cycle: an external driver progressively builds up the electron anisotropy until it exceeds the threshold for whistler instability, at which point whistler waves are excited. The waves then rapidly scatter electrons in the parallel direction, reducing the anisotropy below the threshold, after which the cycle repeats as anisotropy gradually rebuilds. This cyclic pattern of whistler emissions is further validated by experiment-motivated particle-in-cell (PIC) simulations.

The experiments are conducted in the upgraded Large Plasma Device (LAPD) at the Basic Plasma Science Facility of the University of California, Los Angeles (UCLA). LAPD is a linear device with a $20$\,m long plasma column produced by a pulsed DC discharge. Figures \ref{fig:figure1}(a) and \ref{fig:figure1}(b) show the side and end views of the experimental setup. Thermal electrons emitted from a heated cathode are accelerated by a mesh anode and collide with helium gas in the chamber, producing ionized plasma [Figure \ref{fig:figure1}(a)]. The experiments are performed in the quiescent plasma that remains after the DC discharge is switched off (plasma afterglow). The typical afterglow plasma density is $\sim 4 \times 10^{11}$ \si{cm^{-3}}, with an electron temperature of $\sim 0.5$ \si{eV}. A static magnetic mirror, controlled by $10$ sets of independently programmable coil systems surrounding the vacuum chamber, is used to set the magnetic profile in the experiments. The mirror section extends approximately $3.5$ \si{m}, measured between regions where $B = 0.95 B_{\max}$. A $2.45$ \si{GHz} magnetron heats electrons through electron-cyclotron-resonance heating (ECRH). Microwaves propagate in the transverse electric mode with $\hat{E}_{\text {microwaves }} \| \hat{y}$ in the waveguide and become evanescent extraordinary (X-mode) waves as they enter the plasma radially. This process preferentially heat electrons in the perpendicular direction, sustaining an anisotropic population that serves as an energy source for whistler wave excitation \cite{wang2012scattering}. A three-axis high-frequency magnetic probe (commonly referred to as a B-dot probe \cite{everson2009design}) measures the resulting magnetic field fluctuations, positioned outside the magnetic mirror region at $z = 3.2$\,m.

\begin{figure}
\includegraphics[width = 0.8\textwidth]{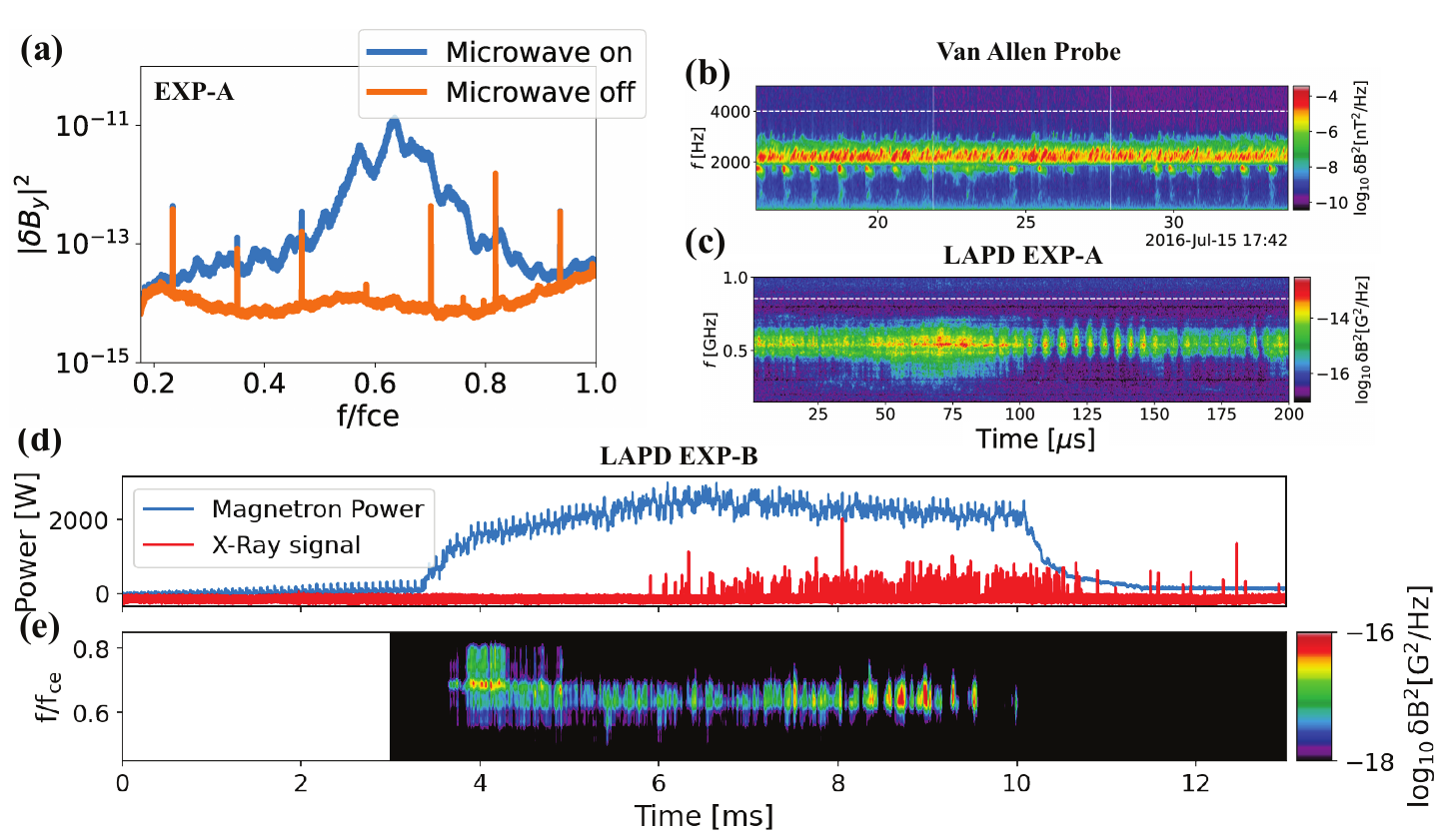}
\caption{\label{fig:fig2} (a) Power spectrum of the magnetic field measured with microwave heating turned on (blue) and off (orange). (b) Van Allen Probe B observation of whistler waves at 17:42:37 UT on 15 July 2016. The white dashed line indicates the electron gyrofrequency $f_{ce} = 4005\, \mathrm{Hz}$. (c) B-dot probe measurement from a single shot in EXP-A. The white dashed line indicates $f_{ce} = 0.84\, \mathrm{GHz}$. (d) Magnetron power (blue) and X-Ray pulses (red) measured by the detector. (e) B-dot probe measurement in EXP-B. Whistler waves are generated after the microwave power is turned on and cease when microwave power is turned off.  }
\end{figure}

Two series of experiments are performed, as indicated in Figures \ref{fig:figure1}(c) and \ref{fig:figure1}(d). In EXP-A, the end of the waveguide is positioned at $x = -31\, \mathrm{cm}$. Different magnetic mirror field profiles are tested (see the Supplemental materials \bibnote{See Supplemental Material for the experimental results under different minimum mirror magnetic field configurations.}), with ECR heating found to be most efficient when the field minimum is $305$\,G ($f_{\mathrm{microwave}} \simeq 3 f_{ce}$). The B-dot probe data is acquired at $3$, $5$, $7$, $9$ and $13$\,ms after microwave pulse initiation, with each shot lasting $200$\,$\mathrm{\mu s}$. Throughout these experiments, the frequency range of the excited waves remains nearly the same. 
Figure \ref{fig:fig2}(a) compares shot-averaged results with and without microwave injection, demonstrating that whistler waves with a peak frequency around $0.63$\,$f_{ce}$ are only excited during microwave heating. The measured maximum wave amplitude is $\sim 5 \, \mathrm{mG}$. Figure \ref{fig:fig2}(c) presents the spectrum of a single shot, where the repetitive excitation of whistler waves is clearly observed. This characteristic is consistently seen in B-dot probe measurements acquired from $3$ to $13$\,ms after the microwave pulse initiation. The repetition period is about $\Delta t_{\mathrm{rep}} \sim 5\,\mu s$, while microwaves heat electrons to the characteristic temperature $m_e (c f_{ce} / f_{pe})^2 = 10$\,keV over a time scale of $\tau_{\mathrm{drive}} \sim 5$\,ms, yielding a ratio $\Delta t_{\mathrm{rep}} / \tau_{\mathrm{drive}} = 10^{-3}$. Such repetitive wave generation is reminiscent of whistler-mode chorus waves observed in the magnetosphere, as illustrated in Figure \ref{fig:fig2}(b). In the magnetosphere, the time scale for electron heating through adiabatic compression during fast plasma injection is approximately $\tau_{\mathrm{drive}} \sim 10$\,minutes \cite{malaspina2018census,zhang2019energy}, while whistler-mode chorus waves exhibit repetition periods of $\Delta t_{\mathrm{rep}} \sim 0.1$-$1$\,s \cite{shue2015local,gao2022observational,kandar2023repetition}, yielding ratios of $\Delta t_{\mathrm{rep}} / \tau_{\mathrm{drive}} = 1/600$-$1/6000$. The similar dimensionless repetition periods in both laboratory and magnetospheric systems suggest that this periodicity is an intrinsic characteristic of such driven systems with respect to whistler anisotropy instability.


In EXP-B, the end of the magnetron horn is positioned at $x = -35$\,cm. Waves are detected when the minimum magnetic field strength ranges between $300$ and $400$\,G. The maximum wave magnitude in EXP-B is $\sim 0.5\,\mathrm{mG}$, which is much smaller compared to EXP-A, indicating lower heating efficiency. Thanks to the high-performance oscilloscope, we continuously measure wave fluctuations throughout the entire heating process in a single shot ($\sim 10$\,ms), capturing the full evolution of the whistler waves excitation process. As microwaves are activated at $3$\,ms, whistler waves begin to emerge, with a clear downshift in the peak power frequency observed from $0.7\,f_{ce}$ to $0.6\,f_{ce}$ at $4$\,ms, where it then stabilizes with repetitive wave elements. A photomultiplier tube (PMT) is deployed outside the vacuum chamber at the magnetic equator to detect X-ray emissions generated when hot electrons strike the metallic surfaces of the plasma chamber. The chamber wall, made of $3/8$\,inch thick stainless steel, blocks X-ray transmission below $\sim 100$\,keV \cite{wang2014enhanced}. X-ray signals intensify $2$--$3$\,ms after microwave initiation, revealing the time required for microwave-driven electron acceleration to reach $100$\,keV, and further indicate the high energy tail is not important in whistler-mode wave excitation. 


These experiments naturally raise two key questions: First, why do the excited whistler waves consistently fall within the frequency range of about $0.6\,f_{ce}$? Second, what mechanism underlies their repetitive nature? To address these questions, we first use linear kinetic theory to calculate the dispersion characteristics of whistler waves. We model the electrons using a single bi-Maxwellian distribution: $f_e\left(v_{\perp}, v_{\parallel}\right)=\frac{1}{(2 \pi)^{\frac{3}{2}} \alpha_{\perp e}^2 \alpha_{\| e}} \exp \left(-\frac{v_{\|}^2}{2 \alpha_{\| e}^2}-\frac{v_{\perp}^2}{2 \alpha_{\perp_e}^2}\right)$, where the thermal velocities $\alpha_{\parallel,\perp}$ are related to the electron temperature through $T_{\parallel,\perp} = m_e \alpha_{\parallel,\perp}^2 $. When the temperature anisotropy is sufficiently large, the distribution becomes unstable to whistler wave growth. To characterize the instability conditions, we define the electron parallel beta as $\beta_{\parallel e}=\frac{2 T_{\| e}}{m_e (c \omega_{ce}/\omega_{pe})^2}$ and fix $\omega_{ce}/\omega_{pe} = 0.15$ ($\omega_{ce} = 2 \pi f_{ce}$ and $\omega_{pe}$ being the angular electron gyrofrequency and plasma frequency, respectively) to match the experimental conditions, allowing the electron distribution to be parameterized by the anisotropy ratio $T_{\perp e}/T_{\parallel e}$ and $\beta_{\parallel e}$ \cite{an2017parameter, yue2016relationship, gary2011whistler}. We then solve the hot plasma dispersion relation using the LEOPARD code \cite{astfalk2017leopard}. 

\begin{figure}
\includegraphics[width =0.8\textwidth]{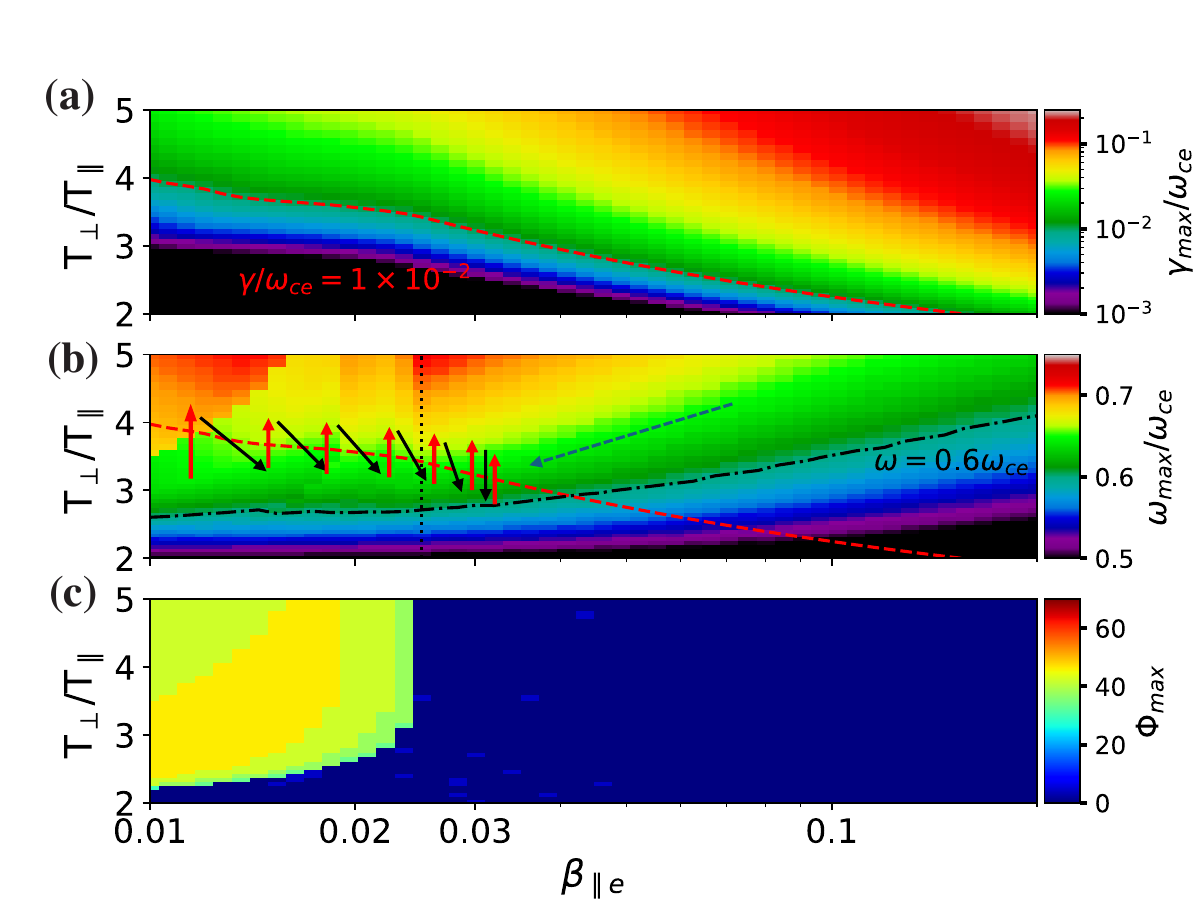}
\caption{\label{fig:fig3} (a) Maximum linear growth rates extracted from wave number space for each plasma state, characterize by $\beta_{\parallel e}$ and $T_{\perp} / T_{\parallel}$. The red dashed line represents $\gamma_{\max} = 10^{-2}\,\omega_{ce}$, indicating the threshold of whistler anisotropy instability. (b) Wave frequency and (c) wave normal angle corresponding to the maximum linear growth rate for each plasma state shown in (a). Solid arrows in (b) sketch the potential evolution path of plasma states in our experiments, while the dashed arrow points to the quasi-steady state with stable whistler wave excitation.}
\end{figure}

The resulting maximum growth rates, along with the corresponding frequency and wave normal angles (WNA), are shown in Figure \ref{fig:fig3}. The assumed instability threshold is plotted as a red dashed line in Figures \ref{fig:fig3}(a) and \ref{fig:fig3}(b). As the external microwave driver progressively builds up the electron anisotropy over time, the anisotropy eventually exceeds the threshold for the whistler anisotropy instability, leading to the excitation of whistler waves. These waves rapidly redistribute the electron distribution, reducing the anisotropy below the instability threshold and suppressing wave growth. The anisotropy then gradually builds up again, perpetuating a cyclical process that sustains wave excitation near the threshold. A potential evolutionary path is illustrated in Figure \ref{fig:fig3}(b) with solid arrows, showing the plasma exceeding the anisotropy threshold, relaxing below the threshold during wave excitation and getting rebuilt again.

As the parallel thermal velocity of the electrons increases, unstable waves begin to grow preferentially along the background magnetic field when $\beta_{\parallel e} > 0.025$, as shown in Figure \ref{fig:fig3}(c) \cite{ma2024nonlinear,yue2016relationship,an2017parameter}. These parallel whistler waves are then detected by the B-dot probe located outside the magnetic mirror, with a measured frequency around $0.7\,f_{ce}$. As the parallel electron temperature continues to rise, the wave frequency downshifts, consistent with EXP-B observations. The results from both EXP-A and EXP-B showing that the frequency remains stable above $0.6\,f_{ce}$ during microwave heating suggest that the parallel electron temperature reaches a quasi-steady state, as indicated by the dashed arrows in Figure \ref{fig:fig3}(b). In this state, a balance is established between microwave heating, wave-driven electron scattering, electron escape through the loss cone, and energy dissipation due to collisions.

\begin{figure*}
\includegraphics[width = \textwidth]{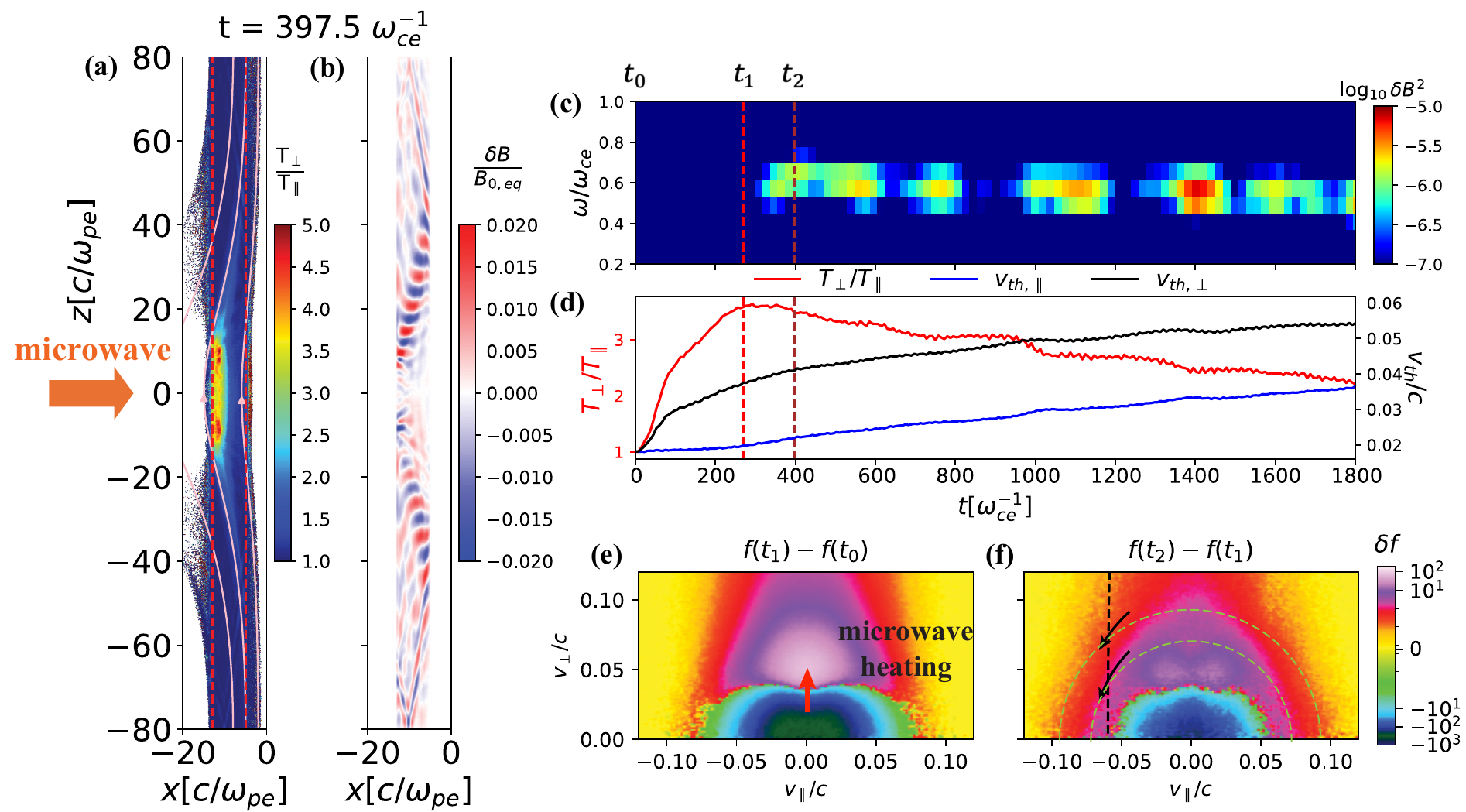}
\caption{\label{fig:4} PIC simulation of whistler wave excitation driven by external microwave heating in a mirror field. (a) Spatial distribution of the temperature anisotropy ratio, $v_{th, \perp}/v_{th, \parallel}$ at $t = 397.5\, \omega_{ce}^{-1}$ after whistler excitation. A linearly polarized microwave ($\hat{E}_{\text {microwaves }} \| \hat{y}$) launched from the $-x$ boundary enters the domain and transitions to an X-mode upon entering the plasma. The initial plasma region from $x = -13$ to $x = -5\, c/\omega_{pe}$ is marked by the red dashed lines. (b) Wave field component $\delta B_{y}$ at $t = 397.5\, \omega_{ce}^{-1}$. (c) Wave spectrogram measured at $(z = 30 \,c/\omega_{pe},\, x = -9 \,c/\omega_{pe})$. (d) Time history of the temperature anisotropy ($T_{\perp} / T_{\parallel}$, red), perpendicular thermal velocity ($v_{th,\perp}$, black), and parallel thermal velocity ($v_{th,\parallel}$, blue), calculated near the field minimum ($x$ ranging from $-12$ to $-11 \,c/\omega_{pe}$ and $z$ ranging from $0$ to $10 \,c/\omega_{pe}$). (e) Phase space density difference between $t_1 = 270\, \omega_{ce}^{-1}$ (before whistler excitation) and $t = 0$, showing perpendicular electron heating by the microwave. (f) Phase space density difference between $t_2 = 397.5\, \omega_{pe}^{-1}$ (after whistler excitation) and $t_1$, illustrating rapid parallel electron heating via cyclotron resonance with whistler waves (black dashed line and arrow) while the microwave remains active.}
\end{figure*}

To gain deeper insight into our experimental findings, we perform 2D3V PIC simulations using the OSIRIS code \cite{fonseca2002osiris} to understand and visualize electron dynamics during whistler wave excitation. In the simulation, a microwave antenna at the $-x$ boundary emits electromagnetic waves with a Gaussian transverse profile (width $\sim 8 \,c/\omega_{pe}$), polarized in the $y$ direction, matching the experimental setup. The microwave frequency is set to $f = 2 f_{ce}$. A static magnetic mirror, similar to the experimental setup, is used with $\omega_{ce} = 0.15 \, \omega_{pe}$ (at the equator) and $\omega_{ce,{\max}} = 0.45 \,\omega_{pe}$. To reduce the computational cost, the simulation domain is assumed to be half the size of the experiment. The initial plasma extends from $x = -13$ to $-5\, d_e$ ($d_e$ being the electron inertial length), while the full simulation domain spans $x = -20$ to $0\, d_e$ and $z = -80$ to $80\, d_e$. The time step is set at $0.02\,\omega_{pe}^{-1}$, and the cell length is $0.03125\,d_e$. Each cell contains 225 particles. The initial thermal velocity of the electrons is $0.018\,c$, corresponding to $167\,\mathrm{eV}$. In the $x$ direction, the particle boundary condition is absorbing, while in the $z$ direction, a thermal bath boundary is used: particles crossing this boundary are reinjected into the simulation with the same initial thermal velocity. Absorbing boundary conditions are applied to all wave fields.

Results show that the electrons are gradually heated in the perpendicular direction due to ECRH (see Figure \ref{fig:4}(d) and Supplementary video \bibnote{See Supplementary video for evolution of electron temperatures in the parallel and perpendicular directions in our PIC simulation.}), increasing their temperature anisotropy. The microwave penetration depth, inferred from the thermal velocity ratio in Figure \ref{fig:4}(a), is $\sim 5\,d_e$, in agreement with theoretical predictions and previous experiments \cite{wang2012scattering}. When the electron temperature anisotropy exceeds the instability threshold, whistler waves are excited, as seen in Figures \ref{fig:4}(b) and \ref{fig:4}(c). Following whistler wave onset, the electron parallel temperature increases as the anisotropy relaxes. Figures \ref{fig:4}(e) and \ref{fig:4}(f) illustrate the evolution of the electron distribution near the equator. Before whistler wave excitation, the microwave primarily heats electrons in the perpendicular direction, as evident from the phase space density difference. Once whistler waves are generated, they rapidly scatter electrons through cyclotron resonance, transferring energy from the perpendicular to the parallel direction, suppressing the instability and leading to a noticeable increase in parallel temperature. This scattering process by whistler waves drives the system toward a more isotropic state.

As the simulation progresses, repetitive whistler wave generation becomes evident. Figure \ref{fig:4}(d) clearly shows that the excitation of each wave element corresponds to a rapid reduction in the temperature anisotropy. Additionally, the increase in parallel electron temperature leads to a downshift in wave frequency [Figure \ref{fig:4}(c)], as well as lowering the instability threshold for temperature anisotropy.
These simulation results confirm that microwave heating effectively energizes electrons and triggers whistler wave excitation in the LAPD parameter space. Furthermore, they provide strong evidence that the experimentally observed repetitive whistler wave excitation is intrinsically linked to the self-regulating nature of whistler anisotropy instability in a driven system. Because the time scale for anisotropy relaxation is significantly shorter than its build-up, the repetition period is primarily controlled by the microwave heating rate, with greater heating power corresponding to a faster repetition rate. The same relation applies to magnetospheric plasma, as the time scale for anisotropy relaxation due to whistler wave generation is $t \simeq 50\,\omega_{ce}^{-1} \simeq 4e^{-3}\,\mathrm{s}$ \cite{an2017parameter,nishimura2002whistler}, which is much smaller than both the repetition period and the electron heating time scale.

In summary, this Letter reports whistler wave generation by electron temperature anisotropy created in a controlled laboratory setting, closely resembling the excitation mechanisms in near-Earth space. Experimental observations, theoretical analysis, and PIC simulations confirm that continuous microwave heating induces electron anisotropy, leading to whistler wave excitation through anisotropy driven instability. This driven system exhibits a repetitive cycle of wave emission and electron anisotropy buildup and relaxation. These experiments open a new avenue for investigating the detailed processes involved in whistler wave excitation, contributing to the understanding of energetic electron dynamics in near-Earth space.

\begin{acknowledgments}
This work was supported by NASA grant NO.~80NSSC20K0917, NSF grant NO.~2108582 and DOE grant NO.~DE-SC0024910. We would like to acknowledge high-performance computing support from Derecho (\url{https://doi.org/10.5065/qx9a-pg09}) provided by NCAR's Computational and Information Systems Laboratory, sponsored by the National Science Foundation \cite{derecho}. The experiment was performed on the Large Plasma Device at the Basic Plasma Science Facility of UCLA, which is a DOE Office of Science, FES collaborative user facility funded by DOE (Award DE-FC02-07ER54918) and the National Science Foundation (Award NSF-PHY 1036140).
\end{acknowledgments}

\appendix

\bibliography{whistler-exp}

\end{document}


\title{Supplementary material to ``Excitation of whistler waves by electron temperature anisotropy in a laboratory plasma"
}

\author{Donglai Ma}
\affiliation{Department of Earth, Planetary, and Space Sciences, University of California, Los Angeles, California 90095, USA}

\author{Xin An}
\affiliation{Department of Earth, Planetary, and Space Sciences, University of California, Los Angeles, California 90095, USA}
\author{Jia Han}
\affiliation{Department of Physics, University of California, Los Angeles, California 90095, USA}
\author{Shreekrishna Tripathi}
\affiliation{Department of Physics, University of California, Los Angeles, California 90095, USA}

\author{Jacob Bortnik}
\affiliation{Department of Atmospheric and Oceanic Sciences, University of California, Los Angeles, California 90095, USA}

\author{Anton V. Artemyev}
\affiliation{Department of Earth, Planetary, and Space Sciences, University of California, Los Angeles, California 90095, USA}

\author{Vassilis Angelopoulos}
\affiliation{Department of Earth, Planetary, and Space Sciences, University of California, Los Angeles, California 90095, USA}

\author{Walter Gekelman}
\affiliation{Department of Physics, University of California, Los Angeles, California 90095, USA}
\author{Patrick Pribyl}
\affiliation{Department of Physics, University of California, Los Angeles, California 90095, USA}
\date{\today}

\maketitle
\section{Abstract}
In this supplementary note, we show the experimental results for different configurations of the minimum mirror magnetic field.
\clearpage

\section{Supplementary Figures}
\begin{figure}[H]
    \centering
    \includegraphics[width=\textwidth]{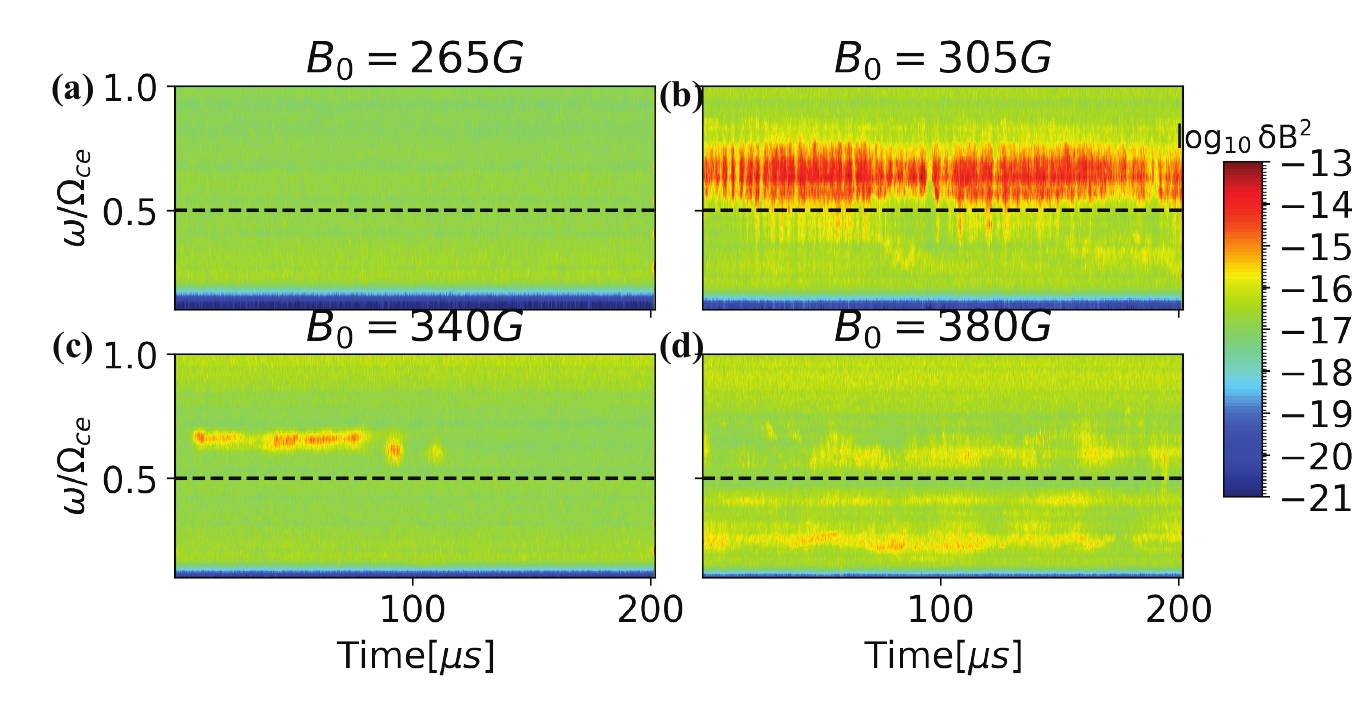}
    \caption{B-dot probe measurements for different minimum mirror magnetic field values. Panels (a)-(d) correspond to $B_0 = 265,305,340,380 \, \mathrm{G}$, respectively. For (a), (c) and (d), the shots with the highest magnetic field fluctuations are selected.}.
    \label{fig:S1}
\end{figure}

Figure \ref{fig:S1} shows four different configurations of the minimum mirror magnetic field, $B_{0}$, in EXP-A. Panels (a)-(d) present B-dot probe measurements for $B_0=265 \,\mathrm{G}$, $305 \,\mathrm{G}$, $340 \,\mathrm{G}$, and $380 \,\mathrm{G}$. The heating efficiency is maximized at $B_0 = 305 \, \mathrm{G}$, where the microwave frequency matches the third harmonic of local electron gyrofrequency.  

\bibliographystyle{apsrev4-2}
\bibliography{whistler-exp}  